\documentclass[apl, pre, twocolumn, amsmath,amssymb]{revtex4-1}
\usepackage{graphicx}
\usepackage{subfigure}
\usepackage{amsfonts}
\usepackage{bm}
\usepackage{dcolumn}
\usepackage{color}

\newcommand{\hh }[1]{ \hat{\bm{#1}} }
\newcommand{\m }[1]{ \mathbf{#1} }

\def\beq{\begin{equation}}
\def\eeq{\end{equation}}
\def\beqa{\begin{eqnarray}}
\def\eeqa{\end{eqnarray}}

\def\Rar{\Rightarrow}

\newcommand{\bfr}{{\bf r}}

\newcommand{\uvec}{{\hh u}}

\newcommand{\bfk}{{\bf k}}

\newcommand{\bfv}{{\bf v}}

\begin{document}

\title{Synchronisation and liquid crystalline order in soft active fluids.}
\date{\today}

\author{M. Leoni$^1$ and T. B. Liverpool $^{1,2}$}

\affiliation{$^1$School of Mathematics, University of Bristol, Clifton, Bristol BS8 1TW, U.K.}
\affiliation{$^2$Isaac Newton Institute for Mathematical Sciences, Cambridge CB3 0EH, U.K.}

\begin{abstract}
We introduce a phenomenological theory  for a new class of soft active fluids, with the ability to synchronise.  
Our theoretical framework describes the macroscopic behaviour of a collection of {\em interacting} anisotropic elements  with cyclic internal dynamics and a periodic phase variable. This system %shows a rich behaviour 
(i) can spontaneously undergo a transition to a state with macroscopic orientational order,  with the elements aligned: a liquid crystal, (ii) attain another broken symmetry state characterised  by synchronisation of their phase variables or (iii) a combination of both types of order. 
We derive the equations describing a spatially homogeneous system and also study the hydrodynamic fluctuations of the soft modes  in some of the ordered states. We find that  synchronisation can promote  the  transition to a state with orientational order;  and vice-versa.
Finally, we provide an explicit microscopic realisation  :  a suspension of micro-swimmers  driven by cyclic strokes.
%; likewise, the alignment 
%can  promote the in-phase synchronisation.
%, as suggested  by   a microscopic realisation   using swimmers  able to synchronise
\end{abstract}
\maketitle

Active materials are composed of self-driven units, active particles, each capable of converting stored or ambient free energy into systematic movement and performing 
work on their surrounding environment~\cite{Joanny-RMP}. 
The interaction of active particles with each other, and with the medium they live in, gives rise to highly correlated collective motion.
Examples include micro-swimmers suspended in a fluid~\cite{Lau09}, motor protein-filaments mixtures, such as those forming  the cell cytoskeleton~\cite{Toner-Tu,Rama,KJJPS,ahmadi}
and  cells forming  tissues~\cite{polarcells}.
They have  non-conventional properties, such as anomalous viscosities~\cite{Sokolov,Rafai}, 
and the ability to self-organise into  ordered states~\cite{concentr-dep},
 with local alignment, 
 forming  patterns~\cite{Riedel}
 and favouring collective transport  on scales larger than individual~\cite{Dombrowski}.  
There is  wide and growing body of theoretical work focussed on investigating the collective dynamics, picturing the individuals as static force-multipoles~\cite{Rama,KJJPS,Saintillan,Aparna} interacting in a fluid
or by generic rules of alignment~\cite{Ginelli}. 
However at the microscopic level, the  dynamics of active individuals  is often time-dependent and cyclic - breaking {\em time-translational invariance}. The effect of this on their collective behaviour is much less well understood.
Recently,  this 
%aspect/property
 has been studied for  swimmers  with {\em identical} cycles by coarse-graining simple dynamical microscopic models~\cite{LL-PRL}.  Whilst this has provided insight, linking the static force multipoles to time averages over the internal cycles, 
it is missing an important property of the system.  In reality the cycles of the individual elements are only identical to within an arbitrary {\em phase} revealing another symmetry.
% that can be broken. 
Therefore the active constituents  typically have the ability to  vary their dynamic cycles  and synchronise their phases via hydrodynamic (or other) interactions thus breaking this {\em phase symmetry}. The effect on the macroscopic behaviour  of active fluids of  this possible broken symmetry   is the subject of this letter.
We note that the subject of hydrodynamic synchronisation of micron-sized oscillators is a %n interesting 
major  topic in  its own right with a long history going back to studies of the coordination of pairs of beating flagella or arrays of cilia beating in a fluid~\cite{GITaylor,GJ07}.
%Driven by experimental developments in imaging 
There have been an 
increasing number of recent experimental studies of systems  investigating the phenomenon  in-vivo~\cite{Polin,Gold}, in vitro~\cite{cilialike} and in minimal artificial systems~\cite{kotar,Qian}.  This has been  paralleled by a recent upsurge in theoretical interest %~\cite{CLJ03,KP04,RS2005,VJ2006,GJ07,NEL2008,UG10,LLPRE2012}.
~\cite{KP04,RS2005,VJ2006,GJ07,NEL2008,UG10,LLPRE2012},
%, a simple model to investigate hydrodynamic synchronisation  had been introduced previously by the authors in the context of oscillators along a line~\cite{LLPRE2012}.
%There  hydrodynamic synchronisation of a large number of  oscillators   is captured by a global phase variable  
%in analogy with the Kuramoto's model for  populations of oscillators~\cite{Ritort}.

In this letter we address the interplay of phase-synchronisation and orientational dynamics in soft active fluids
 considering a minimal  model which consists of active elements
with both an orientation  
and  an internal cycle.  The cycle is characterised by a phase variable which varies slowly with time.  
We define {\em synchronisation} with reference to  the phase dynamics and say it has occurred when the phases of different individuals are locked at  a
fixed phase difference.
 Here we shall consider only  in-phase
 synchronisation, with the phases fixed at the same value, i.e.  the phase difference is zero.
Synchronisation thus viewed is then simply a type of long range order.
Hence, in addition to the usual slow (Nambu-Goldstone) modes 
 describing liquid crystalline fluids and gels~\cite{degennes,KJJPS,ahmadi}, which are associated with broken rotational symmetry,
a theory taking account of synchronisation of the active elements requires another slow mode which is associated with the broken symmetry of the phase and has {\em no classical equilibrium analogue}. 

%Following Landau,
We identify the order parameters of the system (from the possible broken internal symmetries - $U(1) \times SO(d)$ in $d$-dimensions) and the conserved quantities, then 
 obtain the phenomenological equations for their dynamics by including all  the terms allowed by  symmetry.
We  discuss the interplay of synchronisation and orientational dynamics  and find that each type  of order may promote the occurrence of the other. This is supported by  a  particular microscopic realisation of the  system :  a suspension of model  swimmers  driven by internal cycles with varying phase interacting via hydrodynamic interactions. 
%able  to synchronise.
%Finally we comment on the possible equations describing deviations from the bulk and for monopolar systems (such as beating cilia).

\paragraph{Systems where phase and orientation dynamics are uncoupled}
To start, we consider a system with both  orientational  and phase dynamics but where the two are decoupled.
For concreteness, we focus on polar fluids~\cite{KJJPS,ahmadi}, characterised by a mean orientational axis $\m P$ describing states that are not invariant under transformations $\m P \to -\m P$.  A nematic system, whose mean orientational axis $\hh n$ is invariant under transformations $\hh n \to -\hh n$,  can be  dealt with using similar techniques~\cite{MLTBL13}.  
We consider a large number $N$ of microscopic elements, at positions $\bfr_k(t)$,  $k=1, \ldots N$ in a volume $\mathcal{V}$. Each one is characterised by both fluctuating
 orientations $\hh u_k (t)$ and phases $\phi_k(t) \in (0,2 \pi)$. %{\color{red}{which are dynamical and fluctuating  quantities. }}
  The source of fluctuations is in general a combination of thermal and non-thermal effects.
   \begin{figure}[h!]
 \centering
 \includegraphics[width=0.24\textwidth]{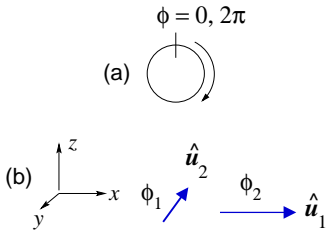}
 \caption{ (a) The  phase variable $\phi$ associated with the internal cycle. Two elements with orientations $\hh u_1$ and $\hh u_2$ and  phases $\phi_1, \phi_2$.  
 To visualise the  phase we schematically represent it  as the length of the arrow.}
 \label{fig:two}
 \end{figure}
A local measure of orientational order is the polar vector,  $\m P( \bfr, t) :=  \langle \frac{1}{N} \sum^N_{k =1} \hh u_k(t) \delta (\bfr - \bfr_k) \rangle$ where $\langle \rangle$ represents an average over the fluctuations.
 Similarly a local measure of synchronisation is the complex-valued quantity~\cite{Ritort,LLPRE2012}
   $\Phi(\bfr, t) := 
 \langle \frac{1}{N} \sum^N_{k =1} e^{i \phi_k(t)} \delta (\bfr - \bfr_k) \rangle$. 
In this letter, we only consider situations where the density,   $\rho := N/\mathcal{V}$  is  constant.

 The average  (bulk)  dynamics of the polar active system is captured by $\m P_0 (t)= \int_r \m P (\bfr,t)$ with an equation of motion~\cite{KJJPS,ahmadi,LL-PRL}, 
 $\partial_t \m P_0 = b \m P_0 - b^{'} \Vert \m P_0 \Vert^2  \m P_0 $.
 The coefficients  $b, b^{'}$ are real valued quantities. $b$  may depend on the density. For swimmers it may also depend on the sign of the average force-dipole~\cite{LL-PRL} of the individuals, while  $b^{'} > 0$  stabilises the magnitude, $P  := \Vert \m P_0  \Vert$ which measures the amount of orientational alignment of the elements, i.e. polar order.
 The
 transition to order is determined by the sign of the coefficient $b$~\cite{ahmadi,LL-PRL}: 
the value $b = 0$  defines the boundary between the orientational disordered phase,
 where  $b < 0$ and the fixed point $ P  = 0$ is stable;
and  polar phase, where  $b > 0$ the  fixed point $ P  = 0$ becomes unstable. 
A generic equation  describing bulk  synchronisation,  $\m \Phi_0 (t)= \int_r \m \Phi (\bfr,t)$  previously considered by us~\cite{LLPRE2012}, is
$\partial_t \Phi_0  = a  \Phi_0  -  a^{'}   |\Phi_0 |^2 \Phi_0 $. 
The coefficients $a, a^{'}$,  are complex valued quantities~\cite{LLPRE2012}. 
 For the following analysis it suffices to focus on their real part $\zeta:= Re[a]$ and $\zeta' := Re[a^{'}]$. 
  $\zeta$ can depend on the density as well as on the isochrony~\footnote{On the limit cycle, oscillations are called non-isochronous if the frequency depends on the amplitude of the oscillations.} of the oscillations~\cite{LLPRE2012}.
  $\zeta' > 0$ is a stabilising term.
%Synchronisation can be analysed by 
Writing $\Phi_0  = M e^{i Q}$,
the magnitude $M$ is a measure of  amount of synchronisation of the elements and quantifies this type of order.
When $\zeta < 0$ the fixed point $M=0$ is stable and 
  the synchronisation  occurs when $\zeta > 0$ and  $M  \ne 0$.
Hence $\zeta = 0$ is the boundary between non-synchronised and synchronised states.
The result of this  uncoupled dynamics can be conveniently  summarised as a phase diagram  in the plane  $(\zeta, b)$, where each phase (disordered, polar,   synchronised $\&$ polar, synchronised) occupies exactly one quadrant, starting from $(\zeta < 0, b < 0)$ in clockwise order.
The interplay of orientational dynamics and synchronisation can shift  some of the boundaries of such a phase diagram,  
 as we will discuss below.

\paragraph{Systems where phase and orientation dynamics  are coupled}
We consider now the average dynamics for this  more general case. %where there is an interplay between the the phases and that of the orientations.
This system is characterised by the density, $\rho$, the polar vector, $\m P_0 (t)$,  and the global phase $\Phi_0 (t)  $ defined above.
 In addition, we must introduce the complex-valued {\em phase-orientation} vector
 $\bm{\Pi}(\bfr,t) := \langle \frac{1}{N} \sum^N_{k = 1} e^{i\phi_k(t)}  \hh u_k(t) \delta ( \bfr - \bfr_k )\rangle$ which encodes the phase-orientation coupling.  As before we can define a bulk value, $\bm{\Pi}_0 (t)= \int_r \bm{\Pi} (\bfr,t)$.  We can understand its physical significance as follows. When  (i) {\em all} the  orientations are identical, $\hh u_k(t) \equiv \hh u(t)$  
 or $(ii)$
 when  {\em all} the phases are locked, $ e^{i\phi_k(t)} =  e^{i\phi(t)}$ $\forall \; k$,  $\bm{\Pi}_0 (t) 
 %= e^{i\phi(t)}   \langle \frac{1}{N} \sum^N_{k = 1}   \hh u_k(t) \rangle 
 \equiv  \Phi_0 (t) \m P_0 (t) $.
 %, where $\Phi_0 (t)$ denotes the global phase variable of the system.
However, when there is partial order  of both of the fields, $\bm{\Pi}_0 (t) 
 \ne  \Phi_0 (t) \m P_0 (t) $. A simple example showing 
that   $\bm{\Pi}_0 $ may be  non-zero,  despite having $\m P_0  = 0$ and $\Phi_0  = 0$
is  a system of 2 particles with  values 
$\hh u_1(t) = -\hh u_2 = \hh  u(t)$ and   $e^{i \phi_1(t)} =  -e^{i \phi_2(t)} =  e^{i \phi(t)} $. 
%Then
%$\bm{\Pi}_0  = \frac{1}{2} \hh u(t) e^{i \Phi_0 (t)}$ so that the magnitude of $\bm{\Pi}_0  \cdot \bm{\Pi}_0 ^*$ has a non-zero value,  despite being $\m P = 0$ and $\Phi_0  = 0$.
The  variables  $\m P_0 (t)$, $\Phi_0 (t)$ and $\bm{\Pi}_0 (t)$ are dynamically coupled.
Equations  of motion can be  obtained by including all the  possible terms  allowed by the symmetries. For simplicity we shall restrict ourselves to the leading order coupling terms, which are quadratic functions.

\begin{figure}[h!]
 \centering
 \includegraphics[width=0.24\textwidth]{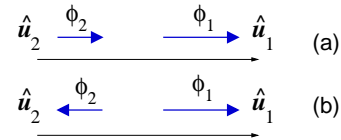}
 \caption{ Elements that are aligned but have different phases. (a) and (b)  contribute differently to $\bm{\Pi}_0  \cdot \m P_0 $. }
 \label{fig:PPi}
 \end{figure}

%{\it Synchronisation dynamics.}
In presence of coupling to orientational dynamics, the equation for  $\Phi_0 $   becomes 
\begin{equation}
\partial_t \Phi_0  = a  \Phi_0  -  a^{'}   |\Phi_0 |^2 \Phi_0  + g \m P_0 \cdot \bm{\Pi}_0   
+  \ldots \label{eq:Phi}
\end{equation}
%where the have neglected the higher order couplings, such as   $\Vert \m P_0  \Vert^2 \Phi_0  $
%and $\Vert \bm{\Pi}_0  \Vert^2 \Phi_0  $. 
%which cannot be derived by considering two-body interactions.
Here % $a, a^{'}$ are defined as above and
 $g$ is  a complex-valued  coefficient.
As eq~(\ref{eq:Phi}) describes a scalar quantity, all its terms can be understood using configurations where the particles are aligned.
Since it is quadratic, the last term can be understood using  2 particles. A finite $\m P_0 $ means  that the configurations
(a) $\hh u_1(t)  = \hh u_2(t) = \hh u(t) $, of fig~\ref{fig:PPi}(a),
and (b) $\hh u_1(t) =-\hh u_2(t) = \hh u(t)  $, of  fig~\ref{fig:PPi}(b), both of which have phases $\phi_1(t)$, $\phi_2(t)$,
should contribute differently to the equation. % Indeed that is confirmed.  
Using the definitions above, in case (a) one finds $\m P_0  \cdot \bm{\Pi}_0  = P_0 ^2 \Phi_0 $; whereas in case (b), $ \m P_0  \cdot \bm{\Pi}_0 = 0$\footnote{If there are correlations such that
 flipping the orientation $\hh u_k \to \hh -\hh u_k$ corresponds to a phase shift $\phi_k \to -\phi_k$ these two cases would be identical. This happens for force monopoles.
}.
%
%{\it Polar dynamics.}
%In presence of interplay with the phase dynamics the equation for 
The dynamics of the polar vector  becomes
 \begin{align}
& \partial_t \m P_0  = b \m P_0  - b^{'} \Vert \m P_0 \Vert^2  \m P_0  
+ h \Phi_0 ^* \bm{\Pi}_0  + h^* \Phi_0  \bm{\Pi}_0 ^*  
  + \ldots  \label{eq:P}
\end{align}
%Here we neglect higher order  terms such as $  |\Phi_0 |^2 \m P_0  $ and $ \Vert \bm{\Pi}_0  \Vert^2  \m P_0  $
  %$b,b^{'}$ are just as defined above and
where $h$ is a complex-valued coefficient.
%The  complex conjugate $h^* \Phi_0  \bm{\Pi}_0 ^*$ is required for the equation to be  real valued.
%
%{\it Complex-valued vector dynamics.}
Finally, we introduce the equation for  the complex polar vector $\bm{\Pi}_0  $ 
 \begin{align}
& \partial_t \bm{\Pi}_0  = q \bm{\Pi}_0  -  q^{'} \Vert \bm{\Pi}_0  \Vert^2  \bm{\Pi}_0   +  m \Phi_0  \m P_0  
+ \ldots  \label{eq:Pi}
 \end{align}
 %where we  neglect higher order terms such as $ |\Phi_0 |^2 \bm{\Pi}_0  $ and $ \Vert \m P_0   \Vert^2   \bm{\Pi}_0 $.
Here the coefficients $q,q^{'},m$  are complex-valued quantities.  $q$ may depend on the density and we consider $Re[q^{'}] > 0 $ providing a stabilising term. 
We can get a simple explanation of the last term in eq~(\ref{eq:P}), coupling $\Phi_0 $, $\bm{\Pi}_0 $, and  the last term in eq~(\ref{eq:Pi}), coupling $\m P_0 $, $\Phi_0 $, 
%in eq~(\ref{eq:Pi})
by  considering  2  particles, with orientations $\hh u_1(t), \hh u_2(t)$ and phases $\phi_1(t), \phi_2(t)$. 
Then using the definition of $\m P_0 $ and $\bm{\Pi}_0 $ and taking the time derivatives, $\partial_{t} \m P_0  = \frac{1}{2}[\dot{\hh u}_1 +\dot{\hh u}_2]$
and $\partial_t \bm{\Pi}_0  = \frac{1}{2} \sum^N_{k=1} e^{i \phi_k} [ \dot{\hh u}_k  + i \dot{\phi}_k \hh u_k ]$,
which combines both the rotational dynamics~\cite{LL-PRL} and the phase dynamics~\cite{LLPRE2012}.
Similarly, $\partial_t \Phi_0  = \frac{i}{2}[\sum_{k=1,2} e^{i \phi_k}   \dot{\phi}_k ]$.
%To see this, we use the  approximation with no noise, where $\bm{\Pi}_0  = \frac{1}{N} \sum^N_{k=1} \hh u_k e^{i \phi_k}$. Then
We note the dynamics  of  orientations and phases will be of the form
$\dot{\phi}_k = \Omega_k + \mathcal{G}^{(k)}(\phi_1 -\phi_2, \hh u_1, \hh u_2)$ and
 $\dot{\hh u}_k = \bm{\mathcal{U}}^{(k)}(\phi_1 -\phi_2, \hh u_1, \hh u_2)$, %where the  scalar and vector functions $ \mathcal{G}^{(k)}$  and $\bm{\mathcal{U}}^{(k)}$ 
 %depending on $\phi_1 -\phi_2$ 
 because of time-translation invariance. 
 The symmetry of the problem suggests an expansion in a  basis of functions given by products of $e^{i (\phi_1-\phi_2)}$ and orthogonal Cartesian tensors constructed from $\hh u^1, \hh u^2$.
The lowest order terms are 
 $\bm{\mathcal{U}}^{(1)} \sim  \hh u_{2} (1 + \lambda_+\cos(\phi_1 -\phi_2) + \lambda_- \sin(\phi_1 -\phi_2)  ) $ 
 where the dynamics of one orientation is determined only by the interactions with the other element.
     $\bm{\mathcal{U}}^{(2)}$ is obtained by exchanging $1 \leftrightarrow 2$. 
 %$\bm{\mathcal{U}}^{(2)} \sim  \hh u_{1} (1 + \lambda_+ \cos(\phi_1 -\phi_2) - \lambda_- \sin(\phi_1 -\phi_2)) $. 
 Similarly $\mathcal{G}^{(1)} \sim  [1 +  (\hh u_1 \cdot  \hh u_2)  ]( 1+\nu_+\cos(\phi_1 -\phi_2) + \nu_- \sin(\phi_1 -\phi_2))$  and
 $\mathcal{G}^{(2)} $ is obtained by exchanging $1 \leftrightarrow 2$.
 %$\sim   [1 +  (\hh u_1 \cdot  \hh u_2)  ](1 + \nu_+\cos(\phi_1 -\phi_2) - \nu_- \sin(\phi_1 -\phi_2) )$. $\lambda_{\pm}, \nu_{\pm}$ are real-valued.
%By homogeneity 
%$\mathcal{T}_{\alpha \beta}, \mathcal{S}_{\alpha \beta}  \propto \delta_{\alpha \beta}$
%and $\mathcal{S}_{\alpha \beta \lambda \nu } \propto (\delta_{\alpha \beta} \delta_{\lambda \nu}  + \delta_{\alpha \lambda} \delta_{\beta \nu} + \delta_{\alpha \nu} \delta_{\beta \lambda})$
%where $\delta_{ij}$ is
% the Kronecker symbol.
%Introducing  the above expressions and 
Inserting these expressions into the equations above and using the definitions of $\m P_0 , \bm{\Pi}_0 $ and $\Phi_0 $,  we get  the last terms on the rhs of eqs.~(\ref{eq:Phi})~(\ref{eq:P}) and~(\ref{eq:Pi}).

\begin{figure}[h!]
 \centering
 \includegraphics[width=0.37\textwidth]{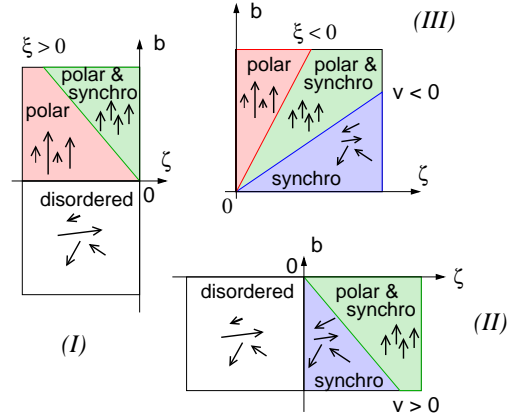}
 \caption{
  Phase diagram, in the plane $(\zeta, b)$: interplay between phase \& orientation dynamics. 
% If phase and orientation dynamics were decoupled, the phase boundaries 
%would simply correspond to the four quadrants of the plane.
% Due to their  interplay   %between synchronisation and orientational dynamics  
 % some of those  boundaries may be shifted.
 We discuss 3 important cases. 
 $(I)$  Polar order promotes the transition to synchronisation
 in region which is forbidden if they are decoupled
  provided    $\xi > 0$. 
  $(II)$ Likewise,  synchronisation  promotes the transition to polar order  in a region which is forbidden to the decoupled system, provided    $v > 0 $. 
  %Transition to polar in this case would occur prior than expected from the study of the uncoupled system.
  In both cases the region of  both synchronisation $\&$ polar order  is widened.
  $(III)$ The opposite  situation arises when  $\xi, v < 0$:
there synchronisation inhibits the transition to polar order; similarly, the  polar order works against synchronisation.
   As a result  %the boundaries are shifted  in such a way that
    the region with both synchronisation and polar order  is reduced compared to the uncoupled system.
   Other intermediate cases are also possible.
 }
 \label{fig:interplay-polar-synchro}
 \end{figure}

%\paragraph{Interplay of  synchronisation and polar order.} 
Instead of analysing the  complex dynamics resulting from 
 the  system of eqs.~(\ref{eq:Phi})~(\ref{eq:P}),~(\ref{eq:Pi}),
we focus here on a particular subset of their state space, analysing the behaviour around the
 fixed points of synchronisation and polar order.
To this end, we eliminate the field $\bm{\Pi}_0 $  in favour of $\m P_0 $ and $\Phi_0 $, 
by setting $\partial_t \bm{\Pi}_0  = 0$ in eq~(\ref{eq:Pi}).   
This generates higher order terms in the equations for $\m P_0 $ and $\Phi_0 $, using
$  \bm{\Pi}_0  = -\frac{ m}{q}  \Phi_0    \m  P_0 $  (and   $  \bm{\Pi}_0 ^* = -\frac{ m^*}{q^*}  \Phi_0 ^*  \m  P_0 $).
 %We then study small deviations  from the fixed points, to gain insight on the interplay of synchronisation and orientational dynamics.
%The microscopic realisation  of swimmers able to synchronise described below indicates that $e $ is small compared to $a$ and $b$ justifying  this procedure a posteriori.
%We consider the simplified problem, obtained by eliminating  the field $\bm{\Pi}_0 $ in favour  of $\m P$ and $\Phi_0 $. In this case 
%Doing so the equations reduce to  
%$\partial_t \Phi_0  = a  \Phi_0  -  a^{'}   |\Phi_0 |^2 \Phi_0    +  s \Vert \m P_0  \Vert^2 \Phi_0   $
%and
%$ \partial_t \m P_0  = b \m P_0  -  b^{'} \Vert \m P_0 \Vert^2  \m P_0  + v  |\Phi_0 |^2 \m P_0  $
 %where $s :=   -g \frac{m}{q} $ and $v : =  -2Re[ h \frac{m}{q} ] $. Here we assume $Re[q] < 0$. 
 We proceed as above, writing $\Phi_0  = M e^{i Q}$ and  $\m P_0  = P  \hh P $
 hence obtain  coupled equations for the magnitudes $M$ (synchronisation) and $P$ (polar order)  
 \begin{align}
 & \partial_t M =  [ \zeta   - \zeta^{'}   M^2    +  \xi   P^2 ] M     \nonumber \\
 & \partial_t  P = [b  - b^{'}  P^2   + v  M^2]  P   \label{eq:M-P}  
\end{align}
where $s :=   -g \frac{m}{q} $ and $v : =  -2Re[ h \frac{m}{q} ] $ and $\xi := Re[s]$. Here we assume $Re[q] < 0$.
%Then there is another pair of equations for the phase $Q$ , slave to them,
%$ \partial_t  Q =  [ \chi   + \chi^{'}   M^2    +  \iota    P^2 ]    $
%and for the director, 
%$\partial_t \hh P = 0  $.
%Here we have further defined  $\chi := Im[a]$,  $\chi^{'} := Im[a^{'}]$  and $\iota := Im[s]$.
%
Setting $\xi = v = 0$, the terms associated with them being higher order than those with  
  $\zeta, \zeta^{'}$ and $b, b^{'}$,
we obtain the unperturbed fixed points of  equations~(\ref{eq:M-P}).
These are: $ (M_0, P_0) = (0 , 0)$, the disordered phase ;  $ (M_0, P_0) = (\tilde{M} , 0) $, the synchronised phase; the polar phase $ (M_0, P_0) = (0 , \tilde{P}) $ and the synchronised $\&$ polar phase $ (M_0, P_0) = (\tilde{M} , \tilde{P}) $. In all these cases, $\tilde{M} := \frac{\zeta}{\zeta^{'}}$ (with $\zeta>0 $) and $\tilde{P} := \frac{b}{b^{'}}$ (with $b>0 $).
%We summarise the result here.  
%The  last fixed point is  stable provided the perturbations $ \xi $ and  $v$ are small enough.
The linearised dynamics of eq~(\ref{eq:M-P}) around the fixed points can be studied by 
considering  small deviations $\delta P := P-P_0$ and $\delta M := M - M_0$.
 In the synchronised phase , $\delta M$ relaxes to zero.
The transition to polar order, however,
  signalled by an instability for $\delta P$,  occurs when $b + \frac{\zeta}{\zeta^{'}} v > 0$.
Remarkably,  synchronisation may promote  polar order which  can occur even for $b<0$,  a region forbidden for the uncoupled system, provided $v > 0$.
An analogous behaviour is seen in the polarised   phase where  $\delta P$ relaxes to zero but  an instability for $\delta M$,
 indicating the synchronisation transition,  occurs 
when $\zeta + \frac{b}{b^{'}} \xi > 0$ and can happen even in the formerly forbidden region $\zeta < 0$, provided $\xi > 0$. 
A  phase diagram is  shown in figure~\ref{fig:interplay-polar-synchro}, which also 
covers the regime when both $\xi , v  < 0$.

\paragraph{Hydrodynamic modes \& stability of bulk states}
Broken symmetry states have slow dynamics of their soft modes.
A full description involves studying $2d-1$ modes in $d$-dimensions so we restrict ourselves to one illustrative case.
A state  with macroscopic fixed polarity $\m P,{\bm \Pi}$ and synchronisation $\m \Phi_0$ (with ${\bm \Pi} \ne \Phi_0 \m P$). We choose w.l.g the phases of $\Phi_0,{\bm \Pi}$ to be zero  and explore the soft modes associated with synchronisation alone, $\m \Phi (\bfr,t) = \m \Phi_0e^{ i \phi (\bfr, t)}$. We consider a collection of nonlinear oscillating force dipoles with frequency $\omega$, amplitude $A$ and friction coefficient $\gamma$ suspended in an incompressible viscous fluid in creeping flow. 
In this oscillating state, the fluid velocity, $\bfv(\bfr,t) = \sum_{n=0}^\infty \bfv^{(n)}(\bfr) \sin ( 2 \omega n  t + 2 \phi_n (\bfr) )$ by the Floquet-Bloch theorem.
%The active elements lead to an additional active stress $\bm{\sigma}^a$ and 
The local phase dynamics is  governed by ($\bfv^{(n)}=0, n\ne2$)%(to lowest order in nonlinearity and gradients)
~\cite{Joanny-RMP,LLPRE2012}
\beqa
&&\eta \nabla^2 \bfv (\bfr,t) - \nabla p =\nabla \cdot \bm\sigma^a \; ; \; \nabla \cdot \bfv= 0 \\
&& {\cal D}_t \phi + P_i \omega_{ij} \Pi_j =  \lambda  P_i u_{ij} \Pi_j + \lambda'  P_i u_{ij} P_j +  D_\phi \nabla^2 \phi 
\eeqa
where ${\cal D}_t g (\bfr,t)=  \partial_t g + (\m v + \beta \m P+ \beta' {\bm \Pi}) \cdot \nabla g $, $\omega_{ij}= \frac12(\partial_i v_j - \partial_j v_i)$ and  $u_{ij}= \frac12(\partial_i v_j + \partial_j v_i)$. The oscillating active stress, $\m \sigma^a_{ij}(t)= P_i P_j \left(\alpha + \frac12 A^2 \rho \gamma \omega \sin  ( 2 \omega t + 2 \phi) \right)$ implies $\phi=\phi_2$, and $\alpha=\bar{\sf F d}$ is the average force dipole~\cite{LLPRE2012}.
An instructive case has   $\m P = P\,  \hh z$ and ${\bm \Pi} = \Pi \, \hh x$. With units s.t. $P=\Pi=1$, a linearized analysis of these equations then reveals modes 
$\phi_\bfk(t) = \int_r \phi(\bfr,t) e^{i \bfk \cdot \bfr}, \bfv_\bfk (t) =  \int_r \bfv^{(2)}(\bfr,t) e^{i \bfk \cdot \bfr}$ which relax as 
$\phi_\bfk(t) = \phi_\bfk(0) e^{-t/\tau_\phi (k) }, \bfv_\bfk(t) = \bfv_\bfk(0) e^{-t/\tau_v (k) }$, where $\tau_v^{-1}(k) = \tau_\phi^{-1} (k) = i \beta k_z + i \beta' k_x + D_\phi k^2 + \Lambda(k)$, with 
\begin{math}
\Lambda(k)= \alpha' \left(  \hat{k}_x \hat k_{z} (1 + \lambda (1-2 \hat{k}_z^2)) + 2\lambda'  \hat k_z^2 (1- \hat{k}_z^2) \right)
\nonumber \end{math}
where $\hat k_i = k_i/|\bfk|$ and $\alpha' = \frac{1}{2\eta} A^2 \rho \gamma \omega$.
The synchronised state has  a  long wavelength instability to propagating phase fluctuations reminiscent of metachronal waves~\cite{Yi03}.
It is interesting that the propagation direction is  intermediate between  ${\bm \Pi}$ and $\m P$, as metachronal waves tend to propagate at an angle 
to the beating direction~\cite{Yi03}.
%$. = \sum_{\alpha=1}^N \left \langle F_\alpha d_\alpha \uvec_\alpha \uvec_\alpha \cdot \nabla  \delta(\bfr-\bfr_\alpha)\right\rangle$
%\paragraph{Nematic system}
% The variables describing a  homogeneous nematic system able to synchronise are: the density $\rho$,   the global phase $\Phi(t)$ defined above. In addition there is  
 %the  nematic orientation tensor (in $D$-dimensions) $\mathbb{S}(t) := \langle \frac{1}{N} \sum^N_{k= 1} [\hh u_k(t) \otimes  \hh u_k(t) -\frac{\mathbb{I}}{D} ] \rangle $,
 %and the complex-valued tensor $\Sigma(t) :=  \langle \frac{1}{N} \sum^N_{k= 1} e^{i \phi_k(t)}  [\hh u_k(t) \otimes  \hh u_k(t) -\frac{\mathbb{I}}{D} ]  \rangle$.
 %The dynamic equations for these quantities can be obtained  using the same approach which led us to  eqs~(\ref{eq:Phi})~(\ref{eq:P})~(\ref{eq:Pi}).

 \paragraph{ Microscopics}
 %{\it Swimmers.}
A microscopic picture of the system is provided by a {\em dilute} collection of micro-swimmers, for simplicity confined to a plane.
A minimal swimmer has an oscillating force dipole in an incompressible, three-dimensional fluid (velocity $\bfv(\bfr,t)$) of viscosity $\eta$ and Re=0.  
%The equations of motion are~\cite{LLPRE2012} 
%\beqa
%\eta \nabla^2 \bfv(\bfr) - \nabla p &=& \sum_{\alpha=1}^N F_\alpha d_\alpha \uvec_\alpha \uvec_\alpha \cdot \nabla  \delta(\bfr-\bfr_\alpha) + \bff_{int} \\
 %\dot{F}_\alpha &=& -\frac{k}{\tau} d_\alpha +\mu \frac{F_\alpha}{\gamma} (1-\sigma d^2_\alpha) + \alpha d^3_\alpha \\
 %\dot{x}_\alpha &=& \bfv(\bfr_\alpha + \uvec_\alpha\frac{d_\alpha}{2}) + F_\alpha / \gamma \\
 %\dot{\uvec}_\alpha &=& \uvec_\alpha \times ( (\nabla \bfv ) \cdot \uvec_\alpha +{\hh \omega}_{int} )
 %\label{eq:force_alpha}
 %\eeqa
%
A concrete example is given by the three-bead swimmer~\cite{NG2004}. 
The swimmer with centre of mass at $\bfr$, oriented along $\hh u$, is made up of  three  collinear spheres, of radius $\mathfrak{a}$, %These are collinear, meaning placed along $\hh u$,  
with coordinates $ x_1,   x_2,    x_3$ and linked together by  extensible links with negligible effect on the fluid.  The spheres are subject to collinear forces $f_1, f_2, f_3$  with $f_1 + f_2 +f_3 =0$ (force-free). %Hydrodynamic interactions among the spheres are described by the Oseen tensor~\cite{Doi}.
The links,  $ L_j := x_{j+1}- x_{j}$, for $j = 1, 2$, have dynamics 
$L_j= l_i + d_j(t)$  where $d_j(t)$ represent the swimming stroke.
 %We further set $l_1 = \kappa l$ and $l_2 = l$.
 % $\kappa > 1 $ describes a swimmer with  positive average force-dipole; $\kappa < 1 $  stems for a negative average force-dipole and
 %$\kappa = 1 $ represent a force-quadrupole~\cite{LL-PRL}.
The  forces and displacements are related by $\dot{d}_i = \frac{1}{\gamma} F_i +\uvec \cdot (\bfv(\bfr,t)-\dot\bfr), \; i =1,2$ %$\dot{d}_1 \approx \frac{1}{\gamma}  F_1$ and $\dot{d}_2 \approx \frac{1}{\gamma} F_2$
with $\gamma = 6 \pi \eta \mathfrak{a}$  and $F_1 = - [ 2 f_1 + f_3]  \equiv {\sf F}$, $F_2 := [  f_1 + 2 f_3]$  ($d_1\equiv{\sf d}$).  %and we have neglected the contribution of the hydrodynamic interactions. 
The force  ${\sf F}$ evolves according to~\cite{LLPRE2012} 
\begin{equation}
 {d {\sf F}}/{dt}  = -{k} {\sf d}/ {\tau}+\mu  (1-\sigma {\sf d}^2) {\sf F}/{\gamma}+ \alpha {\sf d}^3 \; ,
 \label{eq:force}
 \end{equation}
 leading to spontaneous oscillations.
To study the dynamics of ${\sf d}(t)$ we introduce a complex amplitude $A = R e^{i \varphi}$, 
${\sf d}(t) = \frac{1}{2} [A e^{i \omega t} +  A^* e^{- i \omega t}]  \Rar $ 
 $\dot{\sf d}(t) =  \frac{i \omega}{2} [A e^{i \omega t} -  A^* e^{- i \omega t}]$. 
% see~\cite{LLPRE2012}.
  $R, \varphi \in \mathbb{R}^1$ are the amplitude, phase of the oscillations.
%In contrast to ${\sf d}$,  
$d_2(t)$ is a prescribed function of time: 
$d_2(t) = d \sin(\omega t + \varphi)$, and $ \dot{F}_2 \approx \gamma \ddot{d}_2(t) $.
    By averaging over the fast oscillation period $T=2\pi \omega^{-1}$ we obtain an effective description~\cite{LL2010EPL,LL-PRL,LLPRE2012} in terms of the 
    %individual slow 
    %variables, which here are 
    orientation $\hh u$,  phase $\varphi$ and amplitude $R$.
Next we eliminate the amplitude $R$ keeping only the slower phase $\varphi$~\cite{LLPRE2012}  and hence define $\m X \equiv (\hh u, \varphi , \bfr )$. 
We consider $N$ such objects characterised by $\m X_k (t)$. %positions and 
%orientations $\{ \bfr_\alpha, \uvec_\alpha\}$ and internal variables (force, displacement) $\{ {\sf F}_\alpha,  {\sf d}_\alpha\}$. 
To obtain averages we use  the one-particle concentration
$c(\m X, t ) := \langle N^{-1} \sum^N_{k=1} \delta(\m X-\m X_k(t))    \rangle $, the density  of elements %with degrees of freedom with  value $\m X$ 
with $\m X$ at time $t$.
% in the large $N$ limit.
Performing averages over $c$~\cite{ahmadi},  
%Using the identity $\langle N^{-1} \sum^N_{k=1} \mathcal{F}(\m X_k(t)) \rangle 
 %=\int d\m X \mathcal{F}(\m X) c(\m X, t) $ valid for any function $\mathcal{F}$
 we can obtain
   the quantities $\Phi (\bfr,t)$,  $\m P (\bfr,t)$ and $\bm{\Pi}(\bfr, t)$  introduced above. % and  study their dynamics.
For this model % allows a quantitative estimate of  the parameters $a$, $b$, $e$ and $g, h, m$, which 
both $v > 0$ and $\xi > 0$, when $R \gg d$.
%
%{\it Force monopoles.}
% We  consider also a collection of spheres of coordinates $\m x_k := d_k \hh u_k$,  
% that can move in  different directions $\hh u_k$ , each one subject to  its driving force $F_k$ for $k =1, \ldots, N$,.
%     For non-interacting spheres dynamics  is described by $\dot{\m x}_k = F_k \hh u_k$ and by the evolution for  $F_k$, given by eq~(\ref{eq:force}). 
%    Including hydrodynamic interactions the model do not show any direct transition to order.
%%In this case we find that the couplings $a, b, e $ are negative, preventing  a   transition to order.
%However, the system may  synchronise after the alignment is achieved  by some  mechanisms. 

In conclusion,
 we have extended the study of active systems to include collections of 
 % consisting of a collection   
 orientable units  {\em with} an internal cycle  characterised by a single phase variable.
 These show two different types of order: synchronisation where the individual  phases are correlated and liquid crystalline order where their orientations are correlated. 
 We derived phenomenological equations describing the dynamics of a spatially homogeneous mixture with hydrodynamic interactions, 
% Our study is inspired to systems such as suspensions of micro-swimmers or growing tissues of cells where each individuals has cyclic dynamics.
focussing on  the interplay of phase and orientation. %analysing the stability of either the  polar  or the synchronised phase.
Our study reveals that  each type of order  can promote  the transition to a state where both types of order are present,
 in a region of the parameters space that would be inaccessible if the two dynamics were decoupled. 
 This is  supported  by a microscopic model of swimmers able to synchronise.
%  However, to fully make a parallel and better understand the consequences,   one needs to develop the hydrodynamic theory of this system.
From the  theoretical point of view, the symmetry breaking associated with the synchronisation transition presents analogies 
with the physical mechanism yielding the Meissner effect in superconductivity or the Higgs mechanism in particle physics. A natural future direction is a complete description of the coupled  hydrodynamic modes. 
%By pushing the analogy  further,  the  synchronisation transition  may introduce a new length-scale
 %in the hydrodynamic description  
 %may and mitigate the long-wavelength instabilities of the polar vector field predicted for swimmer suspensions~\cite{Rama}.

We thank S. F\"urthauer and S. Ramaswamy for sharing a manuscript on a related topic and acknowledge the support of the EPSRC No. Grant EP/G026440/1.

\bibliography{notes}

\end{document}